\documentclass[aps,nofootinbib,floatfix,amsmath]{revtex4}
\usepackage{epsfig}
\usepackage{graphicx}
\usepackage{multirow}

\begin{document}
\title{Stable heavy pentaquarks in constituent models}
\author{Jean-Marc~Richard} 
\email{j-m.richard@ipnl.in2p3.fr} 
\affiliation{Universit\'e de Lyon, Institut de Physique Nucl\'eaire de Lyon,
IN2P3-CNRS--UCBL,\\ 
4 rue Enrico Fermi, 69622  Villeurbanne, France} 

\author{A.~Valcarce} 
\email{valcarce@usal.es} 
\affiliation{Departamento de F\'\i sica Fundamental and IUFFyM,\\ 
Universidad de Salamanca, E-37008 Salamanca, Spain}

\author{J.~Vijande} 
\email{javier.vijande@uv.es} 
\affiliation{Unidad Mixta de Investigación en Radiofísica e Instrumentación Nuclear en Medicina (IRIMED), \\
Instituto de Investigación Sanitaria La Fe (IIS-La Fe) \\
Universitat de Valencia (UV) and IFIC (UV-CSIC), Valencia, Spain}

\date{\today} 

\begin{abstract}\noindent
It is shown that standard constituent quark models  produce $(\bar c c qqq)$ hidden-charm pentaquarks, 
where $c$ denotes the charmed quark and $q$  a light quark,  which lie below the lowest 
threshold for spontaneous dissociation and thus are stable in the limit where the internal 
$\bar c c$ annihilation is neglected. The binding is a cooperative effect of the 
chromoelectric and chromomagnetic components of the interaction, and it disappears in 
the static limit  with a pure chromoelectric potential. Their wave function contains 
color sextet and color octet configurations for the subsystems and can hardly be 
reduced to a molecular state made of two interacting hadrons. These pentaquark 
states could be searched for in the experiments having discovered or confirmed 
the hidden-charm meson and baryon resonances. 
\end{abstract}

\maketitle 

\section{Introduction}
\label{se:intro}
In recent years, many new hadrons have been discovered in the hidden-charm sector, leading to a 
flurry of theoretical works. For a review, see e.g.,~\cite{Chen:2016qju}. Except for $X(3872)$, 
most states have been seen only in one experiment, or in one type of experiment, e.g., $B$ decay 
or production in $e^+e^-$ collisions, and thus await firm confirmation. Among the recent findings, the two states seen 
by the LHCb collaboration~\cite{Aaij:2015tga} have attracted much attention and inspired many 
interesting studies. Several approaches have 
been proposed to describe the LHCb states, and some of them even anticipated the 
discovery. In the molecular approach, these states consist of coupled $\bar D^* \Sigma_c$ and 
$\bar D^* \Sigma_c^*$, and there are some predictions for strange partners. See, e.g., Refs.~\cite{Roca:2015,Yama:2017}
and references therein. 
A variant is the so-called hadroquarkonium~\cite{Duby:2008}, in which a compact charmonium is trapped 
inside an ordinary hadron. This idea has been tested in lattice simulations~\cite{Albe:2017}. Another 
method relies on QCD sum rules, see, e.g.,~\cite{Wang:2016}. Note that neither lattice simulations nor 
QCD sum rules are fully {\em ab-initio} yet, due to the complexity of the computations, 
and use the guidance of specific models, such as diquarks, to select the operators. 
In the quark model approaches, the attention is often focused on the chromomagnetic part 
of the interaction~\cite{Yuan:2012}. Among the exceptions, we can mention~\cite{Take:2017}.

For sure, this physics has not been exhausted yet. Several experimental searches are limited by 
the small production cross section, and the restriction to specific triggers, e.g., $J/\psi$ in 
the final state, while flavor exotic configurations are also awaited~\cite{Richard:2016eis,Valcarce:2016pew}.
But, as shown  in this letter, even the hidden-charm sector has not yet been fully explored, and some states 
have perhaps escaped our scrutiny. 

Most multiquarks, so far, are resonances, and necessarily involve an interplay between re-scattering effects 
and collective dynamics. Bound states are more easily described with normalizable wave functions, and
when the binding becomes deeper and deeper one expects a transition from a dominant long-range hadron-hadron 
interaction towards a dominance of interquark dynamics.

The constituent quark model has often been used for exploratory studies whose results have been refined 
and confirmed by more rigorous treatments of QCD. For instance, the prediction of flavor-exotic mesons 
$(QQ'\bar q\bar q)$ has been made first by many potential-model calculations and later reinforced
by lattice simulations and QCD sum rules~\cite{Chen:2016qju,Richard:2016eis,Francis:2016hui}.

In this letter, we revisit the hidden-charm configurations $(\bar c c qqq)$, where $c$ is the charmed quark 
and $q=u$ or $d$, using a standard potential model and estimate consistently the mass of the pentaquark 
states and of the ordinary hadrons constituting their dissociation threshold, to identify states which are stable if the internal annihilation of the $\bar c c$ pair is neglected, i.e., lie below their lowest dissociation threshold. The model and the method used 
for the calculations are described in Sec.~\ref{se:model}. Our results are displayed in Sec.~\ref{se:resu}, 
and commented upon in Sec.~\ref{se:disc}. Finally, some other promising configurations are listed in 
Sec.~\ref{se:outl}, which will be the subject of further studies, possibly with an improved modeling of 
the confining interaction. 
\section{The model}
\label{se:model}
We use a simple quark model consisting of non-relativistic kinetic energy and a color-additive interaction 
corresponding to pairwise forces mediated by color-octet exchanges. The validity of this picture 
has been discussed at length in many papers and review articles, and hardly needs further comments. 
The model is used for exploratory studies that could stimulate investigations within more sophisticated pictures. 
To be more specific, we choose the so-called AL1 model of Semay and Silvestre-Brac \cite{Semay:1994ht}, 
which has been already used for multiquark calculations, see, e.g., \cite{Janc:2004qn,Vijande:2009kj}. It reads
\begin{equation}\label{eq:model1}
\begin{gathered}
V(r)  =  -\frac{3}{16}\, \tilde\lambda_i \cdot \tilde\lambda_j 
\left[\lambda\, r - \frac{\kappa}{r}-\Lambda + \frac{V_{SS}(r)}{m_i \, m_j}  \, \vec\sigma_i \cdot \vec\sigma_j \right] \, ,\\
V_{SS}  = \frac{2 \, \pi\, \kappa^\prime}{3} \,\frac{1}{\pi^{3/2}\, r_0^3} \,\exp(- r^2/r_0^2) ~,\quad
 r_0(m_i, m_j)  =  A \left(\frac{2 m_i m_j}{m_i+m_j}\right)^{-B}~,
 \end{gathered}
 \end{equation} 
where
$\lambda=$ 0.1653 GeV$^2$, $\Lambda=$ 0.8321 GeV, $\kappa=$ 0.5069, $\kappa^\prime=$ 1.8609,
$A=$ 1.6553 GeV$^{B-1}$, $B=$ 0.2204, $m_u=m_d=$ 0.315 GeV, and $m_c=$ 1.836 GeV. Here, $\tilde\lambda_i \cdot \tilde\lambda_j$ is a color factor, suitably modified for the quark-antiquark pairs.
We disregard the small three-body term of this model, introduced to fine-tune the baryon masses vs.\ the meson masses.
We also use a variant with slightly different quark masses and parameters 
$\lambda$ and $\kappa$, and a fixed smearing radius for the spin-spin potential,
that has already been utilized to study the six-quark problem~\cite{Vijande:2016nzk}. 

Note that our aim is to show whether the constituent quark model, if taken seriously, 
may lead to binding of some $(\bar c cqqq)$ 
configurations. In quark models, mass differences are usually better 
predicted than the masses themselves. In our case, what is best predicted is 
$m(h)-m(h_1)-m(h_2)$ for a multiquark $h$ with threshold $h_1+h_2$, so changing 
the parameters of the potential will change the masses but barely the binding 
energy of the multiquark $h$.

\begin{figure*}[t]
\hspace*{1cm}\resizebox{12.cm}{17.cm}{\includegraphics{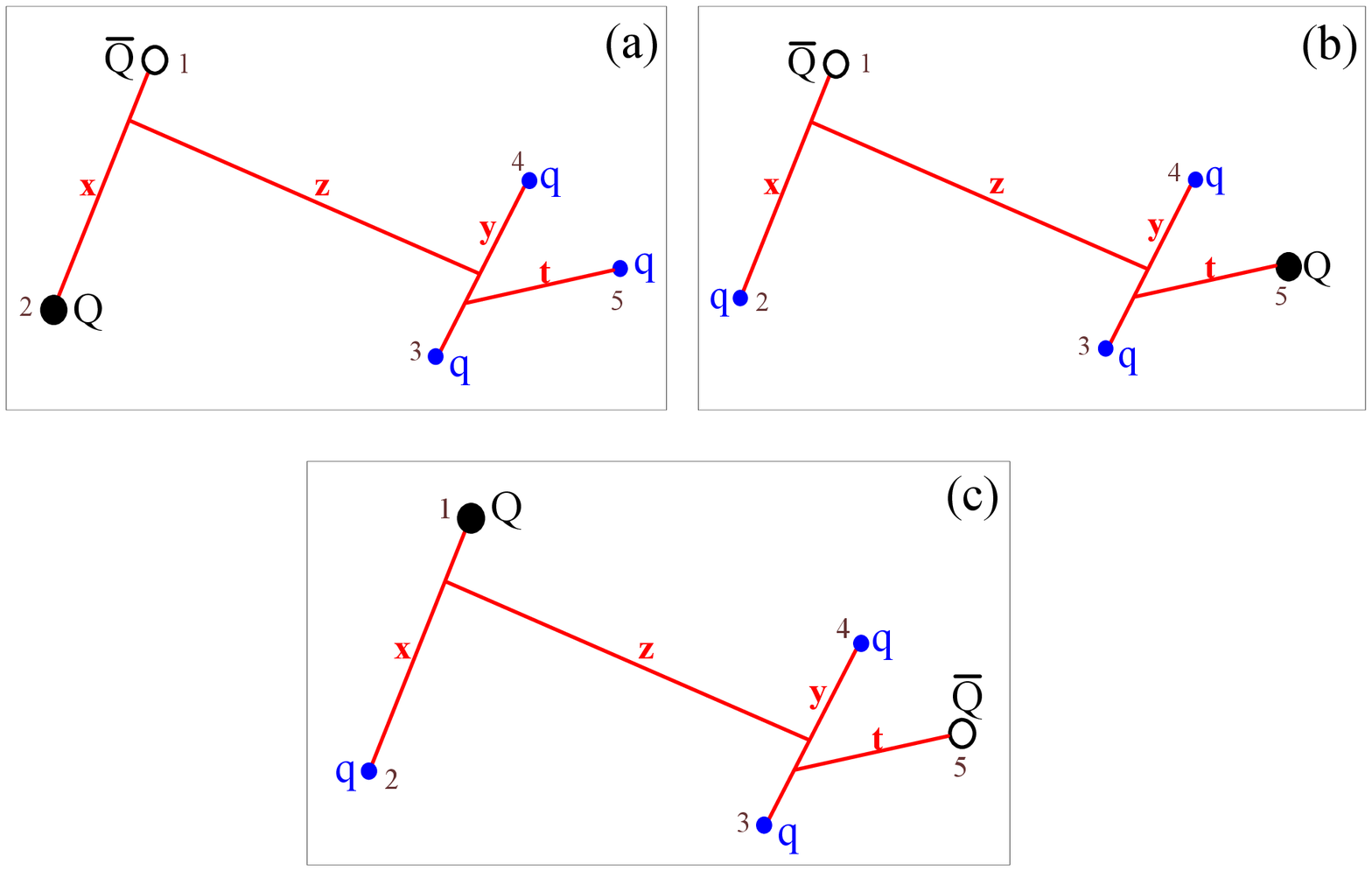}}
\vspace*{-9.0cm}
\caption{The three sets of quark rearrangements and the relative Jacobi coordinates.}
\label{fig:jaco1}
\end{figure*}

To solve the 5-body problem, we introduce the Jacobi coordinates
\begin{equation}\label{eq:jacobi2}
 \begin{gathered}
\vec{x} = \vec{r}_2 - \vec{r}_1 \, , \quad
\vec{y} = \vec{r}_4 - \vec{r}_3  \, , \quad
\vec{t} = \vec{r}_5 - \frac{\vec{r}_3 + \vec{r}_4}{2} \, ,\\
\vec{z}= \frac{\sum_{i=1}^2 m_i \vec{r}_i}{\sum_{i=1}^2 m_i} - \frac{\sum_{i=3}^5 m_i \vec{r}_i}{\sum_{i=3}^5 m_i} \, , \quad
\vec{R} = \frac{\sum_{i=1}^5 m_i \vec{r}_i}{\sum_{i=1}^5 m_i}\, ,
\end{gathered}
\end{equation}
with  different numbering of the $(\bar Q Q qqq)$ constituents.
Namely, three different quark arrangements,  shown in Fig.~\ref{fig:jaco1}, generate 
three sets of Jacobi coordinates.  The arrangements $(a)$ and $(b)$ simulate the asymptotic 
thresholds that contribute to the five-quark state and are explicitly reached when 
the range parameters, of the trial wave function defined below, associated to the 
Jacobi coordinate $\vec z$ vanish. The asymptotic state 
is made of a charmonium and a light baryon for $(a)$ and of an anticharmed meson and a charmed 
baryon for $(b)$. In $(c)$, the thresholds are somewhat hidden and, instead, a confined 
diquark-diquark-antiquark configuration is put forward. Clearly, these choices of Jacobi 
coordinates should in principle lead to the same spectrum. However, the convergence of the 
variational calculation turns out easier if one adopts a quark arrangement rather than another 
one, depending on the pentaquark state. In particular, for each set of quantum
numbers we have used the three Jacobi coordinate arrangements to improve and speed convergence. Note that this strategy is not identical to the one 
promoted by Kamimura {\em et al.}, who astutely mix in the wave functions configurations corresponding 
to different sets of Jacobi coordinates \cite{Hiyama:2003cu}.

In a  simple model such as \eqref{eq:model1}, color, which enters through the $\tilde\lambda_i \cdot \tilde\lambda_j$ factor and the statistics,  is treated as a global degree of freedom, similar to isospin in nuclear physics. 
There are three independent color states for the pentaquarks, which can be chosen for arrangements $(a)$ and $(b)$ as:
\begin{enumerate}
\item $(\bar c c)$ singlet coupled to $(qqq)$ singlet,
\item $(\bar c c)$ octet coupled  to the first $(qqq)$ octet, in which the quarks 3 and 4 are in a $\bar 3$ state,
\item $(\bar c  c)$ octet associated to the second $(qqq)$ octet, in which  the quarks 3 and 4 form a sextet,
\end{enumerate}
and for arrangement $(c)$, as\footnote{This a drawback in the analysis of 
Ref.~\cite{Park:2017jbn} together with the choice of a set of Jacobi coordinates that does not take into account the existence
of particles with different masses.} 
\begin{enumerate}
\item $(Q q)$ in a $\bar 3$ state coupled to the first $(qq\bar Q)$ triplet, in which quarks 3 and 4 are in a $\bar 3$ state,
\item $(Q q)$ in a $\bar 3$ state coupled to the second $(qq \bar Q)$ triplet, in which the quarks 3 and 4 are in a $6$ state,
\item $(Q q)$ sextet associated to the $(qq\bar Q)$ $\bar 6$ state, in which quarks 3 and 4 form a $\bar 3$ state.
\end{enumerate}

The color states are explicitly spelled out in a computer code, using the SU(3) Clebsch-Gordan 
coefficients provided in~\cite{Alex:2010wi} and the associated web site. In this basis the matrix 
elements of the color coefficients $C_{ij}=-3 \,\tilde\lambda_i \cdot \tilde\lambda_j/16$ are 
obvious for the pairs (1,2) and (3,4). For the other pairs, new color states are generated, 
deduced by permutations in the quark sector, and their overlap (crossing matrices) with the initial 
color states are calculated. The same pedestrian method is used for the spin-states, 
to estimate the matrix elements of the spin-color operators $C_{ij}\,\vec\sigma_i\cdot\vec\sigma_j$. 
For a total spin $S=5/2$, there is only one spin state, and thus 3 color-spin states $|\alpha\rangle$. 
For $S=3/2$, there are 4 spin states, and thus 12 $|\alpha\rangle$ color-spin states. For $S=1/2$,  there 
are 5 independent spin configurations, and then 15 color-spin vectors  $|\alpha\rangle$ in the wave function. 

A similar situation is found with the isospin basis (note that for the spin coupling all arrangements are identical,
because we are dealing with five particles of spin $1/2$) that depending on the particle arrangement used the isospin
vectors, and thus their symmetry properties, are different. There are two linearly independent isospin $1/2$ vectors
and one isospin $3/2$ vector. As the interaction is isospin independent, the two allowed
isospin $1/2$ vectors are orthogonal and when the Pauli principle is imposed, the number of vectors
in each arrangement may be different.

For each choice of Jacobi coordinates, the wave function is expanded as
\begin{equation}
 \label{eq:gaussian}
 \Psi=\sum_\alpha \psi_\alpha(\vec x,\vec y,\vec t, \vec z)\,|\alpha\rangle\,,\quad
 \psi_\alpha(\vec x,\vec y,\vec t, \vec z)=\sum_i\gamma_{\alpha,i}\,\exp(-\tilde X^\dagger\cdot A_{\alpha,i}\cdot X/2)~,
\end{equation}
where the $A_{\alpha,i}$ are $4\times 4$ positive-definite matrices
whose elements are the range parameters, and 
$\tilde X^\dagger=\{\vec x,\vec y, \vec t,\vec z\}$. All spatial matrix 
elements can be calculated analytically, using standard techniques of 
Gaussian integration~\cite{zinn2002quantum}. The range parameters 
entering the $A_{\alpha,i}$ are optimized numerically to minimize the variational energy.
No particular Anstaz has been used for these parameters, but the minimization MINUIT code
tries random sets whenever a minimum is reached to check that it is not a local minimum.
The number of generalized Gaussians is increased till the result is converged.
Although we have observed that three are enough to reach convergence, we have pushed the calculations
up to five generalized Gaussians, such that the numerical uncertainty of the variational
calculation is smaller than 1 MeV.  
Note that the symmetry requirements of the orbital wave functions have not been imposed a 
priori. They are restored almost exactly by the minimization procedure. This is another reason why 
different sets of Jacobi coordinates  lead to an improved convergence, as we have checked.
\begin{table}[t]
\begin{tabular}{|c|c|c||c|c|c|}
\hline
\multicolumn{3}{|c||}{Baryons} & \multicolumn{3}{c|}{Mesons} \\ \hline
State                   & Calc.   &  Exp.   & State                   &  Calc.   &  Exp.  \\ \hline
$N$		                  &  0.996  &  0.940  & $D$                     &  1.862   &  1.868  \\
$\Delta$                &  1.307  & 1.232   & $D^*$                   &  2.016   &  2.008  \\
$\Lambda_c$             &  2.292  &  2.286  & $\eta_c$	              &  3.005   &  2.989  \\
$\Sigma_c$              &  2.467  &  2.455  & $J/\psi$                &  3.101   &  3.097  \\
$\Sigma^*_c$            &  2.546  &  2.518  &                         &          &         \\
\hline
\end{tabular}
\caption{\label{tab:ordinary}
Masses (in GeV) of the hadrons entering the thresholds, 
calculated in the potential model of Eq.~(\ref{eq:model1}) 
and compared to the experimental values.}
\end{table}

\section{Results}\label{se:resu}
The masses of the hadrons involved in the thresholds, calculated
with the potential model of Eq.~(\ref{eq:model1}) are compared 
to the experimental ones in Table~\ref{tab:ordinary}.
The thresholds of the channels made out of these hadrons are listed in Table~\ref{tab:threshold}. We 
restrict ourselves to odd overall parity, thus the relative angular momentum between 
the meson and the baryon has to be even, $S$- or $D$-wave in practice. 
\begin{table}[b]
\begin{tabular}{|c|c|c|c|c||c|c|c|c|c|}
\hline
\multicolumn{5}{|c||}{$I=1/2$} & \multicolumn{5}{|c|}{$I=3/2$} \\ \hline
     $J$           &$1/2$&$3/2$&$5/2$&  Mass  & $J$              &$1/2$&$3/2$&$5/2$&  Mass  \\ \hline
$N\eta_c$		       & $S$ & $D$ & $D$ &  4.001 & $\Delta\eta_c$	 & $D$ & $S$ & $D$ &  4.312 \\
$NJ/\psi$          & $S$ & $S$ & $D$ &  4.097 & $D\Sigma_c$      & $S$ & $D$ & $D$ &  4.329 \\
$D\Lambda_c$       & $S$ & $D$ & $D$ &  4.154 & $D\Sigma^*_c$    & $D$ & $S$ & $D$ &  4.408 \\
$D^*\Lambda_c$     & $S$ & $S$ & $D$ &  4.308 & $\Delta J/\psi$  & $S$ & $S$ & $S$ &  4.408 \\
$D\Sigma_c$        & $S$ & $D$ & $D$ &  4.329 & $D^*\Sigma_c$    & $S$ & $S$ & $D$ &  4.483 \\
$D\Sigma^*_c$      & $D$ & $S$ & $D$ &  4.408 & $D^*\Sigma^*_c$  & $S$ & $S$ & $S$ &  4.562 \\
$D^*\Sigma_c$      & $S$ & $S$ & $D$ &  4.483 &                  &     &     &     &        \\
$D^*\Sigma^*_c$    & $S$ & $S$ & $S$ &  4.562 &                  &     &     &     &        \\
\hline
\end{tabular}
\caption{\label{tab:threshold}
Relevant thresholds (in GeV) for $J=1/2,\, 3/2$ and $5/2$ states, 
and angular momentum between the meson and the baryon for
isospin $I=1/2$ and $I=3/2$ states.}
\end{table}

The masses obtained for the pentaquark states are listed in Table~\ref{tab:penta}, together with 
the lowest threshold in relative $S$- and $D$- wave. 
As already mentioned, if a mass is estimated as being larger that the lowest threshold, it just 
means that the state is unbound. Any conclusion about a possible resonance, and the estimate of the width of 
the resonance, would require dedicated further calculations, for instance by applying the method of real scaling \cite{Hiyama:2005cf}. 
\begin{table}[t]
\begin{tabular}{|c|c|c|}
\hline
$(J,I)$  	& $(\bar c c qqq)$     & Lowest threshold \\
\hline
 $(1/2,1/2)$	& 4.077              & 4.001 (S) / 4.408 (D)  \\
 $(3/2,1/2)$	& 4.161              & 4.001 (D) / 4.097 (S)  \\
 $(5/2,1/2)$	& 4.429              & 4.001 (D) / 4.562 (S)  \\
 $(1/2,3/2)$	& 4.077              & 4.312 (D) / 4.329 (S)  \\
 $(3/2,3/2)$	& 4.161              & 4.312 (S) / 4.329 (D)  \\
 $(5/2,3/2)$	& 4.429              & 4.312 (D) / 4.408 (S)  \\
\hline
\end{tabular}
\caption{Variational estimate of the pentaquark masses (in GeV) 
in various spin and isospin channels, compared to the lowest $S$-wave and $D$-wave threshold.}
\label{tab:penta} 
\end{table}

Let us first note the degeneracy between $I=1/2$ and $I=3/2$ states, as could have been
expected a priori due to the isospin independence of the potential model in Eq.~(\ref{eq:model1}),
although the result is not trivial due to the requirements of the Pauli principle.
It is worth to emphasize that the quark arrangement schemes shown in Fig.~\ref{fig:jaco1}
are basic to get the lowest energy. In fact, the isospin $1/2$ states do easily converge to their
lowest energy in arrangement $(a)$ and $(b)$, while isospin $3/2$ states converge better in
arrangement $(b)$. Arrangement $(c)$, that was used to look for explicitly exotic states
with a tightly bound diquark substructure, is not favored by any set of quantum numbers.
It is seen that the lowest state for $(J,I)=(1/2,3/2)$ and $(3/2,3/2)$ are found below 
their lowest $S$-and $D$-wave thresholds: $\Delta \eta_c$ and $D\Sigma_c$. 
In fact they are substantially lower, so that they remain 
stable or metastable if one accounts for the width of the $\Delta$ and considers that the 
actual lowest threshold is $N \pi\eta_c$.
As can be seen in Table~\ref{tab:penta} the vicinity of the two lowest thresholds would
also contribute to enhance the potential attractive character of the interaction for 
these quantum numbers. 
The lowest state for $(J,I)=(5/2,1/2)$ is above its lowest $D$-wave threshold while it is below
the lowest $S$-wave threshold. In this case, as highlighted long ago in Ref.~\cite{Hogaasen:1978jw},
the contribution of color vectors different from the singlet-singlet combination would prevent 
by the centrifugal barrier the tunneling of the quarks to combine in a colorless object,
enhancing in this way the stability of this state.
The other states are found above the lowest $S$-wave threshold.
\section{Discussion}
\label{se:disc}
\begin{figure*}[b]
\resizebox{8.cm}{12.cm}{\includegraphics{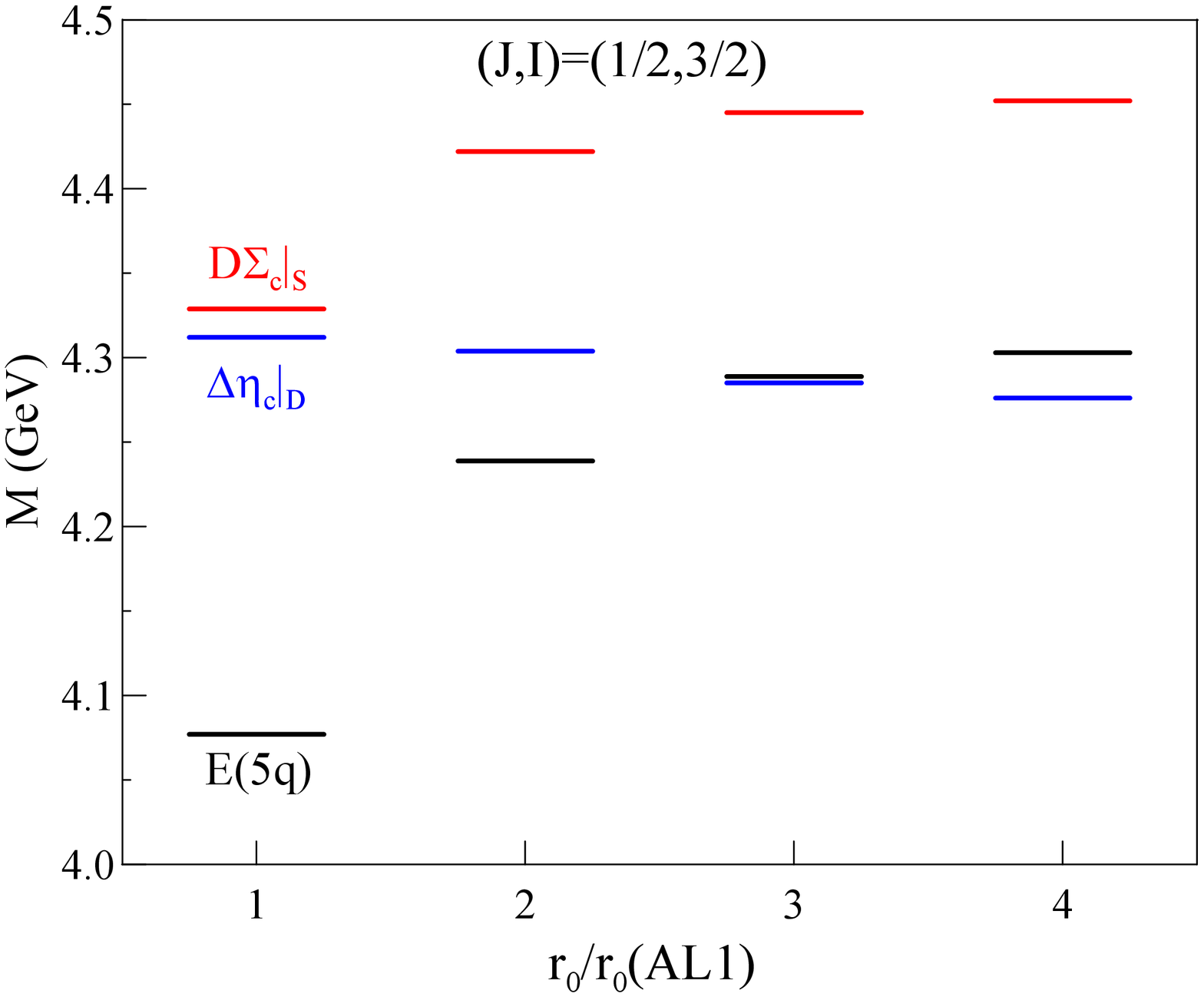}}\hspace*{1.5cm}
\resizebox{8.cm}{12.cm}{\includegraphics{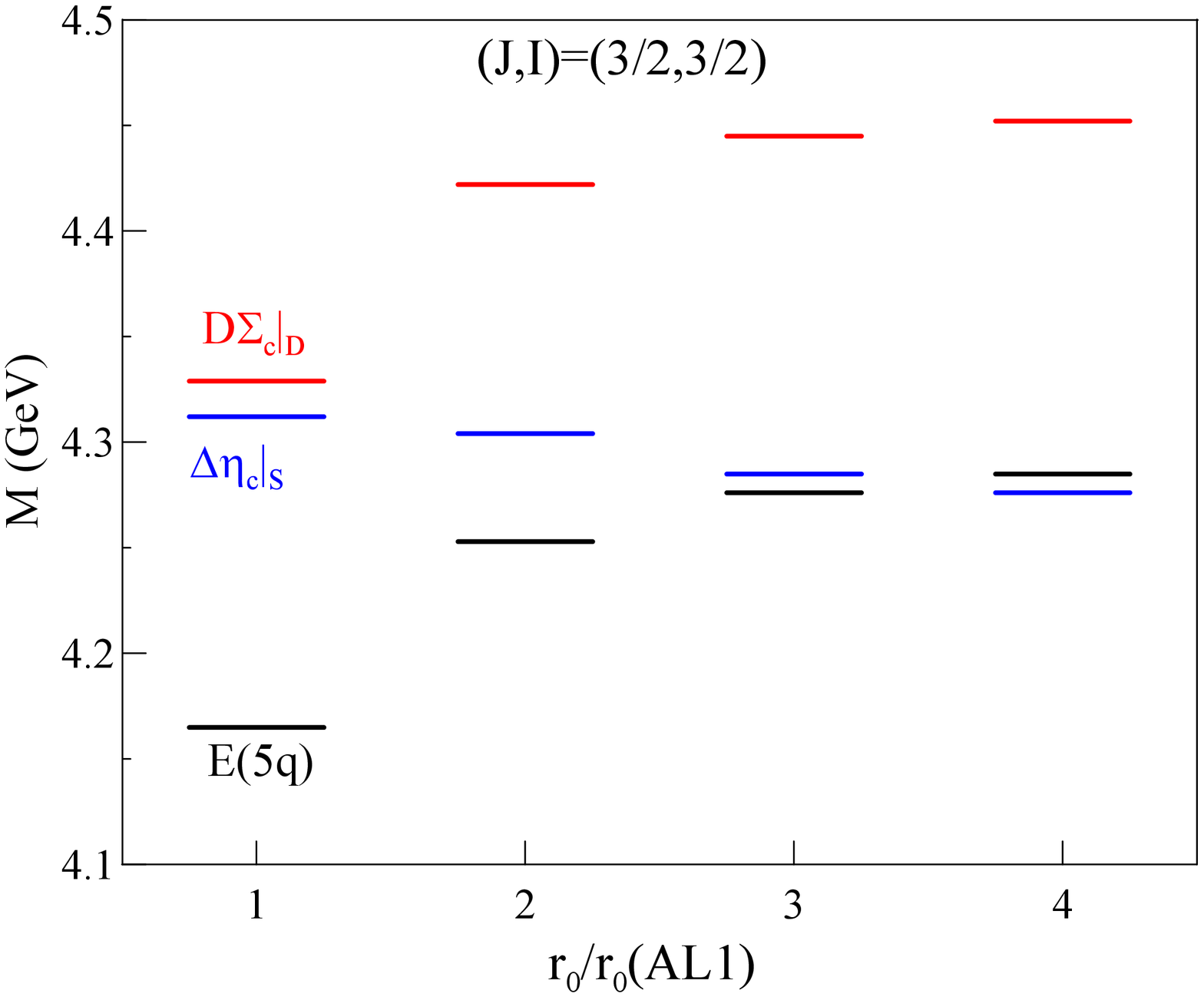}}
\vspace*{-4.0cm}
\caption{Evolution of the masses of the five-quark state and the lowest
$S-$ and $D-$wave thresholds as the smearing parameter of 
the spin-spin interaction is gradually increased.}
\label{fig:change-rz}
\end{figure*}
\begin{figure*}[t]
\resizebox{8.cm}{12.cm}{\includegraphics{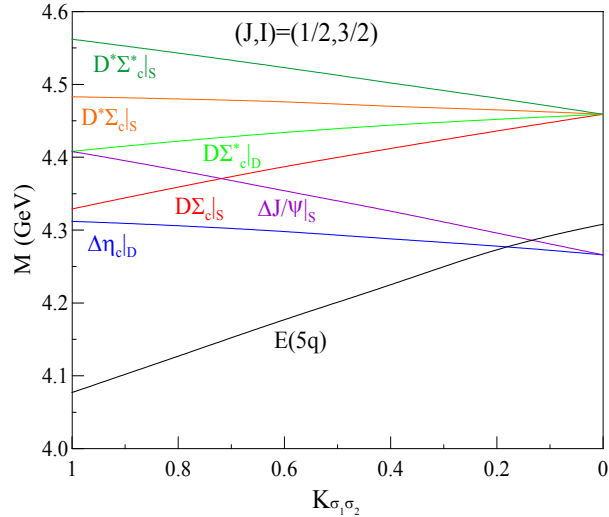}}
\vspace*{-4.0cm}
\caption{Mass of the $(J,I)=(1/2,3/2)$ state and its thresholds when the chromomagnetic 
interaction is decreased by a multiplicative factor $K_{\sigma_1\sigma_2}$.}
\label{fig:spin-spin}
\end{figure*}

The stability of the $(J,I)=(1/2,3/2)$ and $(3/2,3/2)$ ground state with parity $P=-1$ does not depend on 
the detailed values of the parameters of the potential. We have tried several variants, without qualitative 
change of the conclusions. For instance, we have modified the parameters into $\lambda=0.2\,\,$GeV$^2$, 
$\kappa=0.4$ and adopted a simplified spin-spin interaction with a smearing radius frozen at $r_0=1\, $GeV$^{-1}$ 
for all pairs~\cite{Vijande:2016nzk}, and the binding remains. Within the model of Eq.~(\ref{eq:model1}), we have studied the influence 
of the smearing parameter $A$. It is known, indeed, that variational calculations become harder with a superposition 
of long- and short-range potentials. We show in Fig.~\ref{fig:change-rz} the evolution of the masses of 
the five-quark state and the lowest $S$- and $D$-wave thresholds as the smearing parameter of 
the spin-spin interaction is gradually increased. As it can be seen, for the $(J,I)=(1/2,3/2)$ case 
the five-quark state remains below the lowest $S$-wave threshold, although for large values of
$r_0$ it goes above the lowest $D$-wave threshold. For such large values of $r_0$ the chromomagnetic
potential is rather weak, what indicates that such state might appear in nature as a narrow resonance
as explained above~\cite{Hogaasen:1978jw}. Regarding the $(J,I)=(3/2,3/2)$ state, it also survives very
weak chromomagnetic potentials, going above the lowest $S$-wave threshold in the chromoelectric limit.

It is interesting to analyze the role of the spin-spin interaction vs. the spin-independent one.  
At the time of the pioneers in color chemistry \cite{Chan:1978nk}, the eigenvalues of the chromomagnetic 
interaction were calculated for a variety of clusters, including $(qqq)$ in color $8$. See in 
particular~\cite{Hogaasen:1978jw,deSwart:1981upu}. One can see in this  
literature, that the  color-octet $(qqq)$ configuration receives a more 
favorable chromomagnetic energy than the color-singlet $\Delta$ in the threshold. 
However, at the time of color chemistry, attention was paid mainly to states made 
of two colored clusters separated by some angular-momentum barrier, while our study 
deals with an overall $S$-wave multiquark.  

In Fig.~\ref{fig:spin-spin}, it is seen that the binding starts already with a small fraction 
of the spin-spin interaction. This means that there is a favorable interplay of chromoelectric 
and chromomagnetic effects, although the binding disappears in the pure chromoelectric limit.

\section{Outlook}
\label{se:outl}
Our finding opens interesting perspectives, as some similar configurations are also bound in 
this approach. A detailed study will be done in a forthcoming paper. Let us just mention that variations 
in the heavy quark sector generally maintains the binding for spin $1/2$ and isospin $3/2$, as it is 
shown in Table~\ref{tab:heavy-var}.

When moving to the strange sector, the spin-spin 
interaction is magnified due to its $1/(m_im_j)$ dependence and the 
the smaller value of regularization radius $r_0$, as per Eq.~(\ref{eq:model1}),
which is able to maintain a large amount of binding. 

It is interesting to note how the binding energy decreases when
going from the charm to the bottom sector: the
interaction is still attractive, but the degeneracy of the thresholds is lost, and thus the cooperative
effect of the coupled-channels  is less effective. In atomic physics, for instance, the binding of 
the $(m_1^+,m_2^-,m_3^-)$ ions is improved as $m_2\to m_3$, as the  two thresholds made of an atom and 
an isolated charge become degenerate \cite{1977JMP....18.2316H}. When the $H=(uuddss)$ was first calculated 
in the SU(3) limit \cite{Jaffe:1976yi}, the thresholds $\Lambda\Lambda$, $\Sigma\Sigma$ and $N\Xi$ were 
artificially degenerate. It was later shown that with SU(3) breaking, i.e., with the thresholds at 
different masses, the binding is weakened \cite{Oka:1983ku}.
\begin{table}[t]
\begin{tabular}{|cccc|}
\hline
$Q\ Q'$& $(\bar Q Q' uuu)$ &$T_1$ & $T_2$ \\
\hline
$c\ c$  & 4.077 & 4.312  & 4.329\\
$c\ b$  & 7.411 & 7.473  & 7.584\\
$b\ c$  & 7.487 & 7.473  & 7.633 \\
$c\ s$  & 3.094 & 3.269  & 3.103\\
$s\ c$  & 2.877 & 3.269  & 2.958\\
\hline
\end{tabular}
\caption{Variants of the $(\bar Q Q'uuu)$ pentaquark: lowest mass (in GeV) 
compared to the thresholds $T_1=\Delta+(\bar Q Q')$ and $T_2=(\bar Q u)+(Q'uu)$.}
\label{tab:heavy-var}
\end{table}

Except for the deuteron, which is mainly a nucleon-nucleon system, and the doubly-heavy mesons $(QQ\bar q\bar q)$ 
which are not yet seen, there are very few examples of stable multiquarks in the literature. Most experimental 
candidates, and most states predicted in various models, deal with resonances. It is shown in this paper that 
standard potential models generate new bound states in the sector of baryons with hidden-charm. 
If this is confirmed in other approaches, then one of the LHCb pentaquarks could be a kind of 
radial excitation, or, say, a collective excitation of a bound pentaquark and the other
might correspond to the first orbital excitation, with the hope that the 
orbital barrier in the latter case or 
the nodal structure in the former case will somewhat inhibit the decay, and 
make the excited state not too broad in spite of its high mass. Note there are still 
uncertainties about the quantum numbers of the LHCb pentaquarks.
Proposals to study the LHCb pentaquarks in photoproduction of $J/\Psi$ on the proton to verify their 
existence would also help in determining their spin and parity~\cite{Kubarovsky:2015aaa}.

Other states could be bound as well and would deserve some further study:
\begin{itemize}
\item the $(\bar b b qqq)$ analogs, or even better, the $(\bar b c qqq)$ ones, 
which are free of internal annihilation,
\item the naked-anticharm $(\bar c uuds)$ and its analogs obtained by permuting 
$u$, $d$ and $s$ have never been submitted to a detailed 5-body calculation since 
the first study by Gignoux {\it et al.}~\cite{Gignoux:1987cn}, and independently 
by Lipkin~\cite{Lipkin:1987sk},
\item Any $(\bar Q cuds)$ is free of restrictions due to the Pauli principle. Thus 
the confinement can take benefit of the flip-flop interaction and some connected 
diagrams, instead of the color-additive $\tilde\lambda_i \cdot\tilde\lambda_j\,r_{ij}$. 
This provides more attraction~\cite{Vijande:2007ix,Richard:2009rp}. 
\end{itemize}
On the experimental side, it would be interesting to analyze $J/\psi p \pi^-$ final 
state. If the pentaquark lies below the $J/\psi p \pi$ threshold, one could look at 
the isospin violating decay $J/\psi N$ channel, or to any final state corresponding 
to internal $c\bar c$ annihilation, such a  $\mu^+\mu^- p \pi$ or $p\bar p p \pi$ or 
a proton and pions with a $(p,\pi)$ subset in the $\Delta$  
region, and the remaining pions in the charmonium region. 
\section{acknowledgments} 

This work has been partially funded by Ministerio de Econom\'\i a, Industria y Competitividad 
and EU FEDER under Contracts No. FPA2016-77177 and FPA2015-69714-REDT,
by Junta de Castilla y Le\'on under Contract No. SA041U16,
by Generalitat Valenciana PrometeoII/2014/066.
Interesting discussions with Emiko Hiyama, Makoto Oka, and	
Atsushi Hosaka are gratefully acknowledged.

\end{document}